\documentclass{sigchi}

\CopyrightYear{2017}
\setcopyright{acmlicensed}

\toappear{
}

\usepackage{balance}       
\usepackage{graphics}      
\usepackage[T1]{fontenc}   
\usepackage{txfonts}
\usepackage{mathptmx}
\usepackage[pdflang={en-US},pdftex]{hyperref}
\usepackage{color}
\usepackage{booktabs}
\usepackage{textcomp}
\usepackage{caption}
\usepackage{paralist}

\usepackage{subcaption}

\usepackage{microtype}        
\usepackage{ccicons}          

\usepackage{todonotes}

\def\plaintitle{PhyShare: Sharing Physical Interaction in Virtual Reality}

\def\emptyauthor{}
\def\plainkeywords{Virtual Reality; Haptic User Interfaces; Robots;}

\makeatletter
\def\url@leostyle{%
  \@ifundefined{selectfont}{
    \def\UrlFont{\sf}
  }{
    \def\UrlFont{\small\bf\ttfamily}
  }}
\makeatother
\def\pprw{8.5in}
\def\pprh{11in}

\definecolor{linkColor}{RGB}{6,125,233}
\hypersetup{%
  pdftitle={\plaintitle},
  pdfauthor={\emptyauthor},
  pdfkeywords={\plainkeywords},
  pdfdisplaydoctitle=true, 
  bookmarksnumbered,
  pdfstartview={FitH},
  colorlinks,
  citecolor=black,
  filecolor=black,
  linkcolor=black,
  urlcolor=linkColor,
  breaklinks=true,
  hypertexnames=false
}

\begin{document}

\title{\plaintitle}

\numberofauthors{3}
\author{%
  \alignauthor{Zhenyi He\\
    \affaddr{60 5th Ave, Office 342}\\
    \affaddr{New York, NY 10011}\\
    \email{zh719@nyu.edu}}\\
  \alignauthor{Fengyuan Zhu\\
    \affaddr{60 5th Ave, Office 342}\\
    \affaddr{New York, NY 10011}\\
    \email{zhufyaxel@gmail.com}}\\
  \alignauthor{Ken Perlin\\
    \affaddr{60 5th Ave, Office 342}\\
    \affaddr{New York, NY 10011}\\
    \email{ken.perlin@gmail.com}}\\
}

\teaser{ 
  \centering 
  \captionsetup{justification=centering} 
  \includegraphics[width=\textwidth]{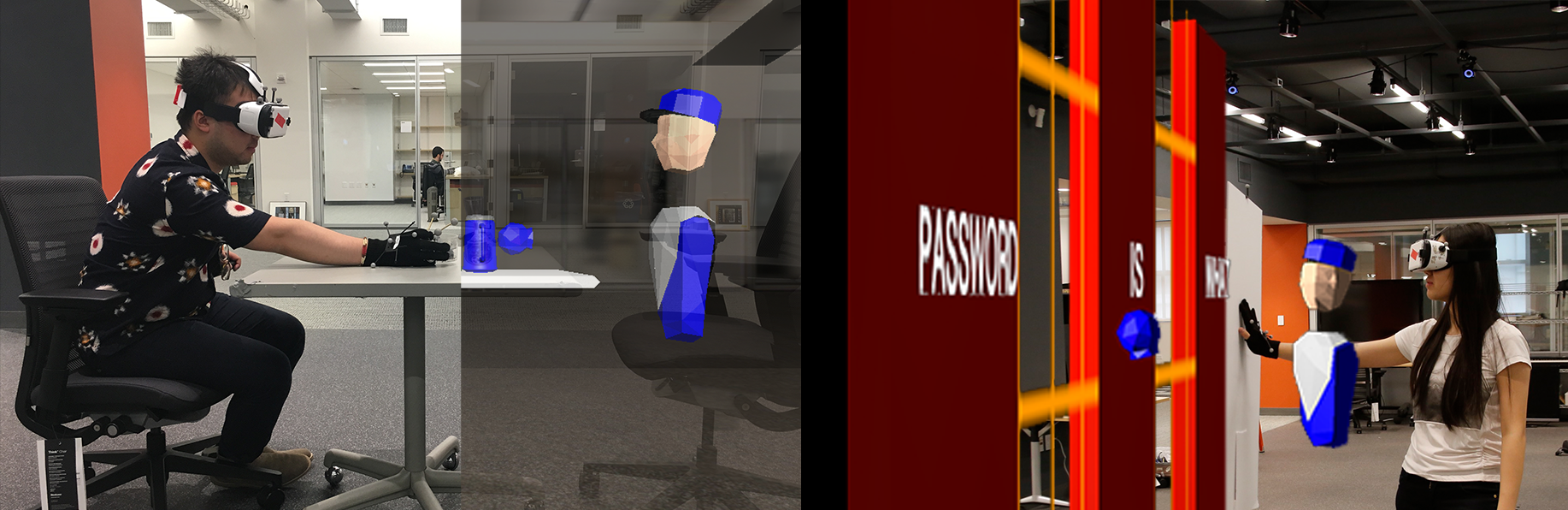} 
  \caption{Sharing Physical Interaction When Clinking the Mugs and Pushing the Wall} 
  \label{fig:teaser} 
}

\maketitle
\begin{abstract}

Virtual reality has recently been gaining wide adoption.
Yet the absence of effective haptic
feedback in virtual reality scenarios can strongly detract from
the suspension of disbelief needed to bridge the virtual and
physical worlds.
PhyShare proposes a new haptic user interface based on actuated robots.
Because participants do not directly observe these robotic proxies, multiple
mappings between physical robots and virtual proxies can be supported.
PhyShare bots can act either as directly touchable
objects or invisible carriers of physical objects, depending on the scenario.
They also support
distributed collaboration, allowing remotely located
VR collaborators to share the same physical feedback.
A preliminary user study evaluates PhyShare for
user awareness, preference, acknowledgement, reliability
and related factors.

\end{abstract}
\category{H.5.m.}{Information Interfaces and Presentation}{Miscellaneous}
\category{H.5.2.}{Information Interfaces and Presentation}{User Interfaces}


\section{Introduction}
This work aims to add to the growing field of
"Robotic Graphics", described by William A. McNeely \cite{mcneely1993robotic} as
"robots simulating the feel of an object and graphics displays
simulating its appearance".


Several significant steps have recently been made towards William A. McNeely's concept, particularly through research on passive objects like MAI Painting Brush\cite{otsuki2010mai,sugihara2011mai}, NormalTouch\cite{benko2016normaltouch}, 
human actuated systems like TurkDeck\cite{cheng2015turkdeck} 
and actuated or wheeled robots such as Zooids\cite{le2016zooids}, CirculaFloor\cite{iwata2005circulafloor}, Snake Charmer\cite{araujo2016snake}, and Tangible Bots\cite{pedersen2011tangible}.

However, current systems suffer from a number of limitations.
First, passive objects and human actuated systems only support
static haptic feedback, which can be insufficient when interacting with a dynamic
environment.
On the other hand, actuated systems in real or in augmented reality (AR) environments aim to control
movable props that correspond to particular virtual objects. However, current implementations do not support dynamic mapping between physical props and their virtual proxies, and might therefore
require large numbers of actuated objects\cite{le2016zooids}, with a corresponding
increase in system complexity.

In addition, 
\cite{leithinger2014physical,richter2007remote,riedenklau2012integrated} which allow remote collaboration based on
distributed actuated tangible objects, provide only a
limited sense of visual feedback rather than a fully immersive experience.
Likewise, many such systems require expensive hardware to function, and are thus are primarily meant to be operated in special laboratory environments rather than embedded in
the physical world.

Our research focuses on variable mappings between virtual
objects in a VR world and their physical robotic proxies,
multiple modes of physical manipulation, and a solution for distributed
collaboration based on wheeled robots (see Figure \ref{fig:teaser}). PhyShare is a hardware and software system: a wheeled robot can behave either as a haptic proxy
or a invisible carrier of haptic proxies,
thereby supporting both
direct manipulation and the illusion of telekinesis
in a shareable virtual world.

PhyShare utilizes the Holojam multi-user VR framework\footnote{\url{https://holojamvr.com/}}
integrated with wheeled robots (m3pi robot\footnote{\url{https://www.pololu.com}},
iRobot\footnote{\url{http://store.irobot.com/default/create-programmable/}}), and the OptiTrack motion tracking system.


Because all interaction takes place in a virtual reality
environment, PhyShare can
choose what to display and what to hide. Therefore, its
mobile robots can
provide haptic feedback when touched, and move
in a way that is invisible to the user
at other times. This "invisibility" allows much
room for optimization in path planning of PhyShare robotic proxies.
For example, if two remote collaborators
are using PhyShare to play tic-tac-toe in VR,
each user can see the actual movement of the remotely
located controller of their opponent.
Meanwhile, in their own physical space, a haptic proxy of
that controller is invisibly moving on the desk
in a path that can vary from that actual movement,
as long as it arrives at the correct location
in time for the local user to touch it.

In summary, our contributions are:
\begin{compactitem}
\item Multiple mapping designs between physical props and virtual proxies to expand the possibilities of robot assistants,
\item Multiple manipulation methods for controlling actuated robots, and
\item A new interaction approach supporting haptic feedback for distributed collaboration in virtual reality.
\end{compactitem}


\section{Related Work}
Our work touches on several research areas, including physical objects in VR and AR, tabletop tangible user interfaces (TUIs) and distributed collaboration.

\subsection{Physical Objects in VR and AR}
Physical objects have the ability to offer tactile feedback for immersive
experiences.
While tactile feedback through stimulating skin receptors alone
is relatively straightforward and continues to enjoy ongoing development,
a number of VR and AR researchers have focused on use of physical
objects for tactile feedback in recent years.

\subsubsection{Passive Objects}
SkyAnchor focuses on attaching virtual images to physical objects that can be moved by the user\cite{kajita2016skyanchor}.
Liao et al. implemented a telemanipulation system that simulates force feedback when controlling a
remote robot manipulator\cite{liao2000force}.
In both cases, the tangible objects in the scene
are passive, only moving when moved by the user.
This approach is not sufficient for general
simulation of dynamic environments.a

\subsubsection{Robotic Shape Display}

As described in Robotic Graphics\cite{mcneely1993robotic}, Robotic Shape Display(RSD) described a scenario in which actuated robots provide feedback when users come
into contact with a virtual desktop. This scenario requires the robot to be ready to move at any time to meet the user's touch.

Some recent work has implemented this idea. 
TurkDeck\cite{cheng2015turkdeck} simulates a wide range of physical objects and haptic effects by using human actuators. While good for prototyping, this approach has
clear limitations in terms of scalability.  A scalable approach
requires a technology that can move without
human operators.


NormalTouch and TextureTouch\cite{benko2016normaltouch} offer haptic shape rendering in VR, using very limited space to simulate feedback
of different surfaces of various objects.
Snake Charmer\cite{araujo2016snake} offers a robotic arm that moves based on its user's position and provides corresponding haptic feedback of different virtual proxies with different textures and temperatures.
Our work shifts the focus to promoting a variety of
manipulation methods between
user and objects. Therefore we focus on direct manipulation,
remote synchronization, and creating the illusion
of telekinesis.

\subsubsection{Roboxels}
Roboxels (a contraction of robotic volume elements), are able to dynamically
configure themselves into a desired shape and size.
Some recent work extends this concept.
CirculaFloor\cite{iwata2005circulafloor} provides the illusion of
a floor of infinite extent. It uses movable floor elements,
taking advantage of the inability of a VR user to see the
actual movement of these floor elements.
Our work also exploits the invisibility of a physical robot
to a person who is immersed in VR,
but takes this concept in a different direction.

\subsection{Tabletop TUI and Swarm UI}
Tangible user interfaces (TUIs) allow their users to manipulate physical
objects that either embody virtual data or act as handles
for virtual data\cite{richter2007remote}.
TUIs can assume several different
forms, including passive sensors\cite{otsuki2010mai,sugihara2011mai,choi2016wolverine} and actuated pieces\cite{araujo2016snake}. 
Table TUIs incorporate physical objects moving on
a flat surface as input\cite{everitt2003two}. 
These objects are used to
simulate autonomous physical objects\cite{brave1998tangible,leithinger2014physical,Pedersen:2011:TBI:1978942.1979384,riedenklau2012integrated,rosenfeld2004physical}.

Tangible Bots\cite{pedersen2011tangible} use wheeled robots as input objects
on a tabletop rear-projection display.
When the user directly manipulates Bots, the Bots in turn
react with movement according to what the user has done.
Tangible Bits\cite{ishii2008tangible,ishii1997tangible} is general framework proposed by Hiroshi Ishiii to bridge the gap
between cyberspace and the physical environment by incorporating active tangible elements into the interface. 
Our work extends this idea of combining the physical world with
digital behavior in immersive VR.

Zooids\cite{le2016zooids} provides a large number of actuated
robots that behave as both input and output. Since
Zooids merge the characters of controller and haptic display, users
can perform manipulations more freely.
A large number of
robots are required in Zooids because the display
requires relatively high resolution. This is not as necessary
in VR environments, because the same
haptic proxy can be used to stand in for different
virtual objects.
In our work, we can use one haptic feedback object as
the physical proxy for multiple virtual objects.



\subsection{Distributed Collaboration}
Some work offers different views for different users\cite{follmer2013inform,gauglitz2014world,leithinger2014physical,leithinger2015shape,sra2016asymmetric,sra2016resolving}.
Among these, \cite{gauglitz2014world} provides only spatial annotations for local users as guidance.
Other work shares the same
environment without haptic feedback, including Holoportation\cite{orts2016holoportation} and MetaSpace\cite{sra2015metaspace}, in which all interactions are established
as belonging either to
local physical objects or remote virtual objects.
InForm\cite{follmer2013inform,leithinger2014physical,leithinger2015shape} introduces an approach to physical
telepresence that includes capturing and remotely rendering the shape
of objects in shared workspaces. Local actuated movables
behave both as input by manipulation and output by shape
changing. This approach is limited by the disparity between
visual and haptic feedback for remote users.

PSyBench\cite{brave1998tangible} first suggested the possibility of distributing TUI. The approach was to virtually share the same objects when users are remote
using Synchronized Distributed Physical Objects. A similar idea called Planar Manipulator Display\cite{rosenfeld2004physical}  was developed to
interact with movable physical objects. A furniture
arrangement task was implemented for this bidirectional interaction. 

Tangible Active Objects\cite{riedenklau2012integrated} extends the idea by adding
audio feedback. This research was based around the constraint of What You Can See is What You Can Feel\cite{yokokohji1996you}. Our own work adapts this core concept and puts it into VR to open up more possibilities, including
multiple mappings between what the user can see (virtual proxies)
and what the user can feel (physical objects).

Some work mentions different mapping possibilities in distributed collaboration\cite{Reilly2011,reilly2010twinspace,richter2007remote}.
TwinSpace\cite{Reilly2011,reilly2010twinspace} extends
the idea of the "office of the future" \cite{raskar1998office} by presenting a framework for collaborative cross-reality. It supports
multiple mixed reality clients that can have different configurations.

One inspiration for our approach of dynamic mapping
was \cite{richter2007remote}, which implemented a remote active tangible interaction
based on a furniture placement task. This work mentions that
a robot could represent different pieces of furniture at different
points in time, likening this to changing costumes. In our research,
we extend this idea of changeable representation and identity to VR.

\section{Hardware and Software Design}

\subsection{Hardware}

\begin{figure}[h]
\begin{subfigure}{.24\textwidth}
  \centering\captionsetup{width = .85\linewidth}%
  \includegraphics[width=0.9\linewidth]{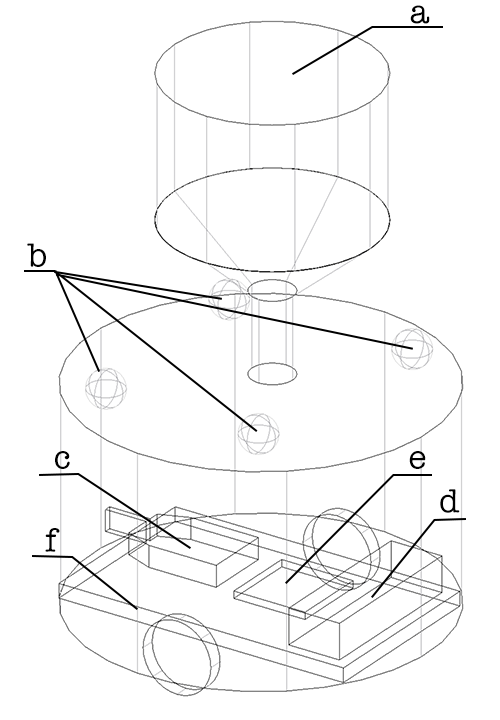}
  \caption{Tabletop Robot Configuration
  \\\hspace{\textwidth}
  a: physical objects, b: rigid body marker, c: XBee wireless module, d: battery, e: micro control board, f: pololu m3pi robot with mbed socket}
  \label{fig:gw:m}
\end{subfigure}%
\begin{subfigure}{.24\textwidth}
  \centering\captionsetup{width = .85\linewidth}%
  \includegraphics[width=0.9\linewidth]{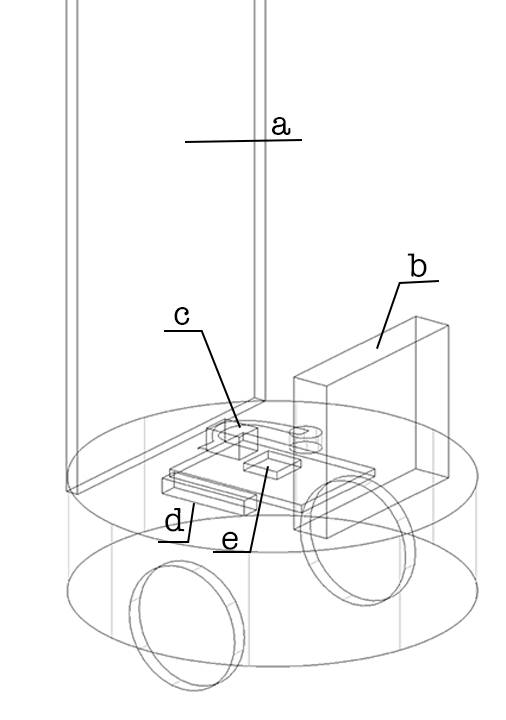}
  \caption{Floor Based Robot Configuration
  \\\hspace{\textwidth}
  a: wall, b: counter weight, c: serial communication module, d: battery, e: raspberry pi}
  \label{fig:hw:i}
\end{subfigure}%
\caption{Hardware Configuration}
\label{fig:hardware}
\end{figure}

We use an OptiTrack for position and rotation tracking.
Rigid body markers are attached to each object to be tracked, including the head-mounted display (HMD) on each user and gloves on each user's hands.
For the HMD, we use a GearVR with Samsung Galaxy Note 4\footnote{\url{http://www.samsung.com/global/galaxy/gear-vr/}},
and use sensor fusion between the GearVR's IMU and the OptiTrack.
For the robots, we use small Pololu m3pi robots equipped with mbed
sockets\footnote{\url{https://www.pololu.com/product/2151}} and iRobot Create 2 as actuated wheeled robots.

There are two different robots used in our system,
one for small tabletop manipulation and one for
large floor based movement.

For the tabletop robot,
the m3pi robot
can function as an interaction object itself--an invisible holder for carrying an interaction object (see figure~\ref{fig:hardware}).
We equip each m3pi robot with an XBee module for communication.
We pair another XBee board to send commands to each m3pi
robot and receive their replies.

The larger floor based robot is based on an
iRobot Create 2, to which a rigid "wall" is affixed
(see figure~\ref{fig:hw:i}).
For communication, we add a raspberry pi\footnote{\url{https://www.raspberrypi.org/}} onto the iRobot to receive commands through WiFi.

\subsection{Software}

\begin{figure}[h]
\includegraphics[width=0.45\textwidth]{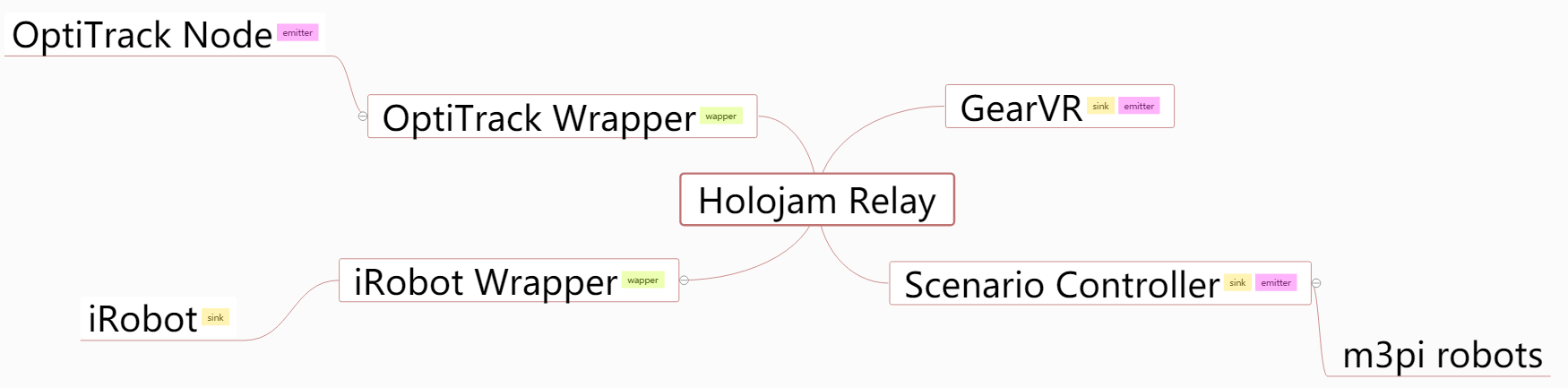}
\caption{networking architecture based on Holojam}
\label{fig:arch}
\end{figure}

To connect our robots together and synchronize their actions with multiple users in virtual reality, we use Holojam, a combined networking and content development framework for building location-based multiplayer VR experiences.
Holojam provides a standard protocol for two-way, low-latency wireless communication, which is flexible enough to use for any kind of device on the network, including our robots.
Essentially, Holojam acts as the central relay node of a low latency local Wifi network, while other nodes can behave as either emitter, sink or both. Each client receives updates from the Holojam relay node (see in figure\ref{fig:arch}).

\section{Design Principles and Challenges}

\subsection{Multiple Mapping}
One challenge of physically representing a virtual environment is the method used to map virtual objects to their physical counterparts (or vice versa). In non-VR TUI systems, there is generally a direct mapping between what users see and what they feel. 
In tangible VR systems, there is more room to experiment with the relationship between the real and virtual environments. 
Conventionally, virtual objects have no physical representation and can only be controlled indirectly or through the use of a standard input device (such as a game controller). Also, the set of physically controlled objects that are visually represented in VR have traditionally been limited to the input devices themselves. In contrast, our system provides three different mapping mechanisms to augment the relationship between physical and virtual representation, in order to explore how each mapping affects user experience and interaction design.

\begin{figure}[h]
\begin{subfigure}{.16\textwidth}
  \centering
  \includegraphics[width=0.9\linewidth]{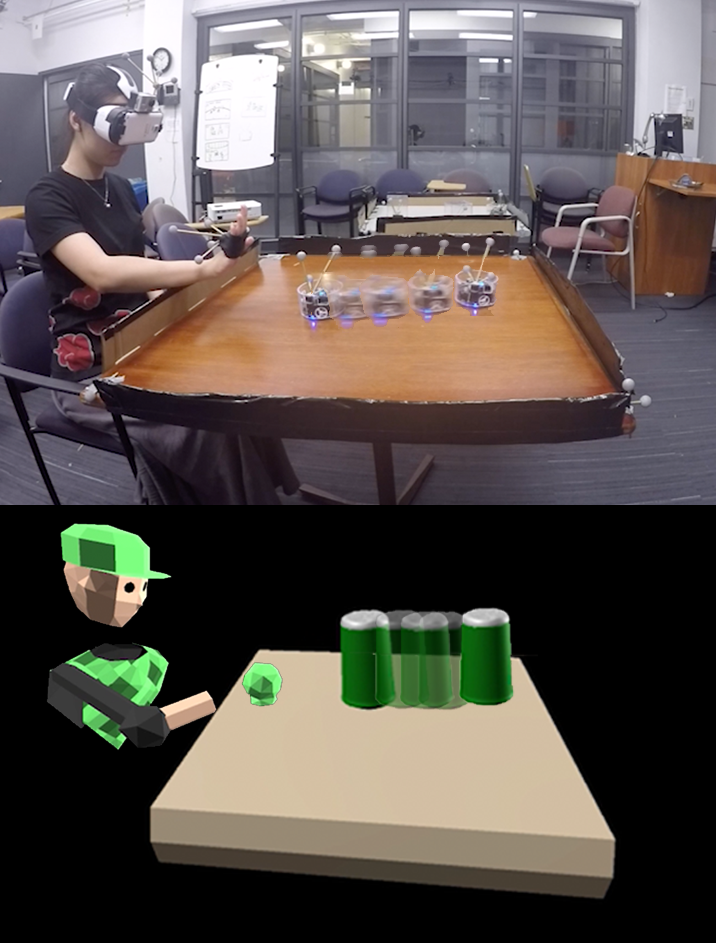}
  \caption{telekinesis}
  \label{fig:usecase:tel}
\end{subfigure}%
\begin{subfigure}{.16\textwidth}
  \centering
  \includegraphics[width=0.9\linewidth]{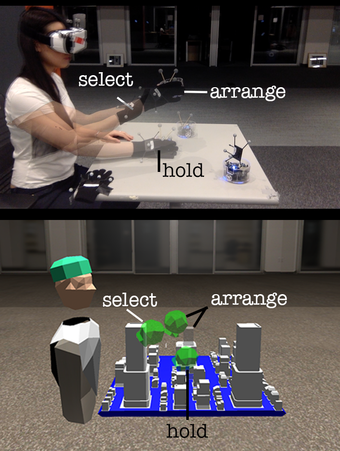}
  \caption{city builder}
  \label{fig:usecase:cb}
\end{subfigure}%
\begin{subfigure}{.16\textwidth}
  \centering
  \includegraphics[width=0.9\linewidth]{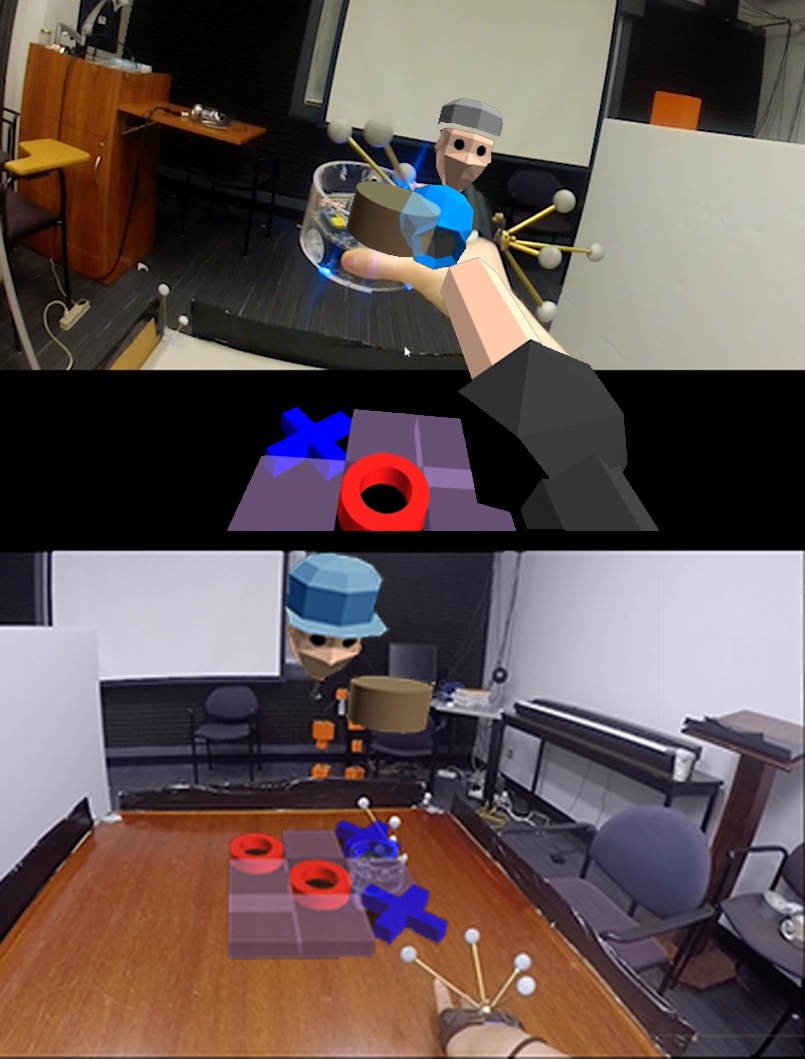}
  \caption{tic-tac-toe}
  \label{fig:usecase:ttt}
\end{subfigure}
\caption{Use Cases}
\label{fig:usecase}
\end{figure}

\textit{One-to-One Mapping} \\
This is the standard and most straightforward option. When users interact with a virtual object in the scene, they simultaneously interact with a physical proxy at the same location.
In the 'telekinesis' use case (see figure~\ref{fig:usecase:tel}), the m3pi robot is represented by the virtual mug.

\textit{One-to-Many mapping} \\
Various constraints, including cost and space, limit the capability of maintaining a physical counterpart for every one of a large number of virtual objects. When fewer proxies are available than virtual objects, one of the total available proxies could represent a given virtual object as required. For example, a user with a disordered desk in VR may want to reorganize all of their virtual items. In each instance of the user reaching to interact with a virtual item, the single (or nearest, if there are multiple) proxy would relocate to the position of the target item, standing by for pickup or additional interactions. The virtual experience is seamless, while 
in the physical world a small number of proxies move into position as needed to maintain that seamless illusion.
In the 'city builder' use case (see figure~\ref{fig:usecase:cb}), a proxy is fitted behind-the-scenes to the virtual building which is visually nearest to a given user's hand. In this implementation, the movement and position of the robots is completely invisible to the user.

In the 'escape the room' use case, the iRobot carries a physical wall which moves with the user to provide haptic feedback. The virtual wall that the user perceives in VR is much longer than the touchable wall section that represents it in the physical world (see in figure~\ref{fig:teaser}).

\textit{Many-to-One mapping} \\
When multiple proxies represent one virtual object, we define the mapping as "many-to-one." This is useful for remote-space applications: A virtual object could exist in the shared environment, which could then be manipulated by multiple users via their local proxies.

The 'clink the mugs' use case simulates the effect of two users striking their mugs together while in separate physical locations (see in Figure \ref{fig:teaser}). In each local environment, the user's mug's position is governed by a non-robotic tracked physical object, and the force of the strike is simulated via a synchronized moving proxy.


\subsection{Multiple Manipulation Methods}

Our system supports several methods of interacting with virtual proxies via physical objects.


We support direct (one-to-one) manipulation, for comparison purposes.
We also employ a custom gesture recognition technique, enabling users to command the position of a target using hand motions.
Utilizing a simple configuration of tracked markers that reports each user's wrist position and orientation via the OptiTrack system, users can push a nearby(proximate) object across the table via a pushing gesture, pull a distant object towards them with a pulling gesture, or motion in another direction to slide the object towards another part of the table.

\subsection{Retargeting in Physical Distributed Collaboration}
Inspired by Synchronized Distributed Physical Objects\cite{brave1998tangible}, we extend the idea and adapt it to a VR environment,
using a novel retargeting\cite{azmandian2016haptic} system. In the 'Tic-Tac-Toe' use case, (figure~\ref{fig:usecase:ttt}), the virtual object does not necessarily represent the physical location of the robot.
While the user is handling the object, remote users see this exact movement. During this operation, the remote robot invisibly moves to the predicted location where the user will set down the virtual object.


For the Tic-Tac-Toe experience, we added "artificial" latency to improve the user experience when synchronizing remote actions, since the robotic proxy takes time to move into position. 
Given the speed of the m3pi robot and the size of the tabletop workspace, we concluded that 1.5 seconds is ideal. 
Movements by the local user of their proxy object are visually delayed in the remote collaborator's view to give the remotely located robot sufficient time to catch up with the local user's actions. This smooths the haptic operations considerably.

In our user study, we test the perception of this latency, with positive results. 
When we asked users directly whether they had experienced any latency, the answer was generally no (see details below).

\section{Preliminary User Study}
The goals of our study were (1) \textit{Necessity}. How do users feel about remote physical interaction? (2) \textit{Awareness of latency}. How much latency do users perceive when interacting remotely? (3) \textit{Preference}. Which manipulation approach does the user prefer? (4) \textit{Awareness of remote location of the other participant}. Does the user feel that their opponent is sitting at the same physical table, or does it feel as though their opponent is remotely located?

Sixteen participants took part in our study, three female. Ages ranged from twenty-two to fifty-two, and the median age was twenty-six. All participants were right-handed. Seventy-five percent had experienced VR prior to the study.

\begin{figure}[h]
\begin{subfigure}{.16\textwidth}
  \centering
  \includegraphics[width=0.9\linewidth]{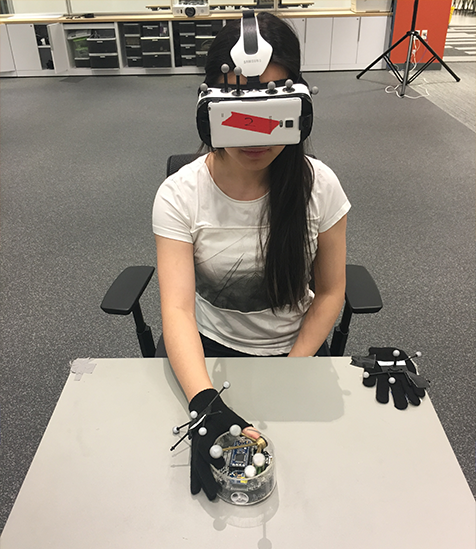}
  \caption{physical tic-tac-toe which move controller to make the move}
  \label{fig:exp:1}
\end{subfigure}%
\begin{subfigure}{.16\textwidth}
  \centering
  \includegraphics[width=0.9\linewidth]{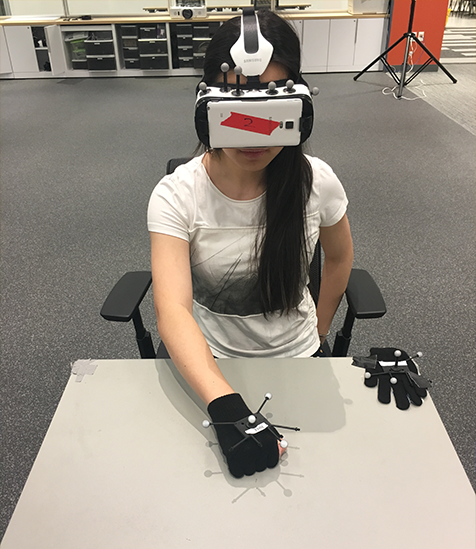}
  \caption{pure tic-tac-toe which use only hand to make the move}
  \label{fig:exp:2}
\end{subfigure}%
\begin{subfigure}{.16\textwidth}
  \centering
  \includegraphics[width=0.9\linewidth]{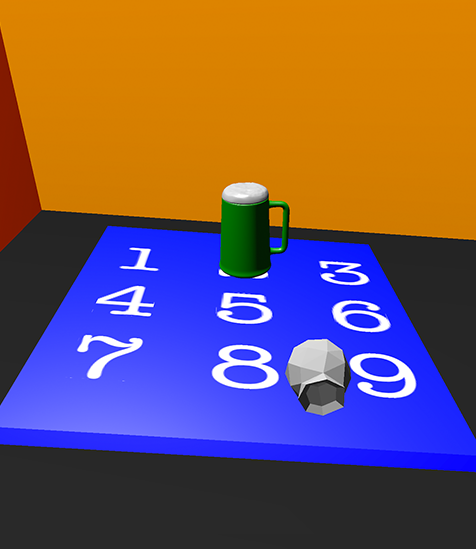}
  \caption{table split by 9 areas which mug is in area 2 for example}
  \label{fig:exp:3}
\end{subfigure}
\caption{Experiment Sketch}
\label{fig:exp}
\end{figure}

\subsection{Experiment Design}


We designed two tests for our user study.
The first was a recreation of the classic board game, \textit{Tic-Tac-Toe}, in VR.
We utilized a "controller" object, allowing players to select a tile for each game step. When one player is holding the controller and considering their next move, the other player can observe this movement in their view as well.
We included a version of \textit{Tic-Tac-Toe} that was purely virtual (no haptics) for comparison (figure ~\ref{fig:exp:1} and ~\ref{fig:exp:2}).

In the second experiment, "Telekinesis" (figure ~\ref{fig:exp:3}), we split the virtual table into nine parts, sequentially placing a target in each one of the subsections. Its purpose was to measure users' ability to learn our gesture system and to use it to command a target at various locations.
Participants were first given up to two minutes to learn the gestures, then we observed their command choice (including non-gestural physical interaction) for each target position.

\subsection{Procedure}
We first explained to participants the purpose of the study. Before each task, the interaction techniques involved were described.

Completion time and error rate were logged for each experiment. In a questionnaire, the participants rated their interaction using four questions from QUIS\cite{chin1988development}: difficult/easy, dull/stimulating, never/always, too slow/fast enough. The questionnaire also included additional questions, using a five-step scale.

\begin{figure}[h]
\includegraphics[width=0.45\textwidth]{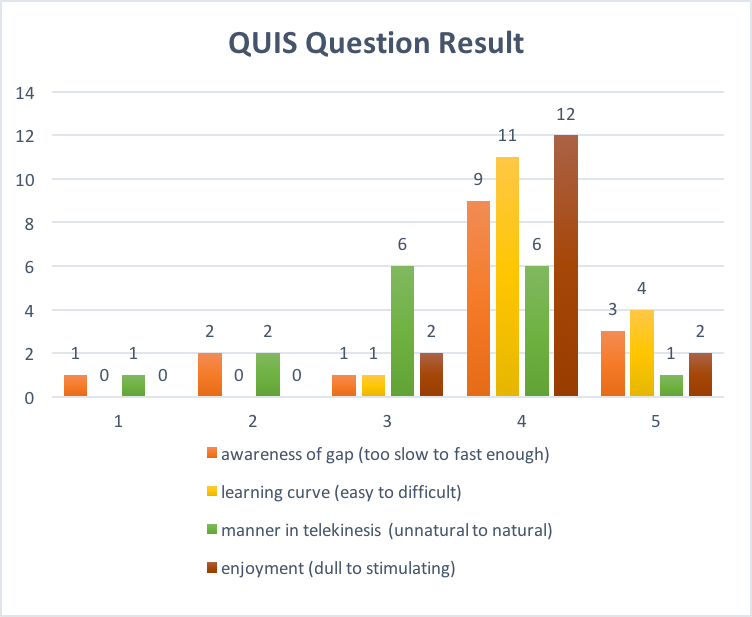}
\caption{QUIS Question Results}
\label{fig:res1}
\end{figure}

\subsection{Results}
The results from our QUIS questions were positive (see figure~\ref{fig:res1}).
For gap awareness (M = 3.93, SD = 1.13), 12 out of 16 (75.0\%) felt the experience was fast enough. Regarding the learning curve of system (M = 4.47, SD = 0.54), 15 out of 16 (93.8\%) thought it was easy to learn.
For the telekinesis experiment (M = 3.47, SD = 1.00), 7 out of 16 (43.8\%) thought the interaction was straight-forward.
Although this result is not high, keep in mind that only 6 out of 16 (37.5\%) scored a 3, and (M = 4.27, SD = 0.52), 14 out of 16 (87.5\%) enjoyed the interaction.

Regarding \textit{necessity}, 10 out of 16 (62.5\%) preferred to play Tic-Tac-Toe with a physical controller (figure~\ref{fig:res3}). Regarding \textit{acknowledgement of physical placement}, 14 out of 16 (87.5\%) felt they were playing on the same table rather than performing a remote task (figure~\ref{fig:res3}).

\begin{figure}[h]
\includegraphics[width=0.45\textwidth]{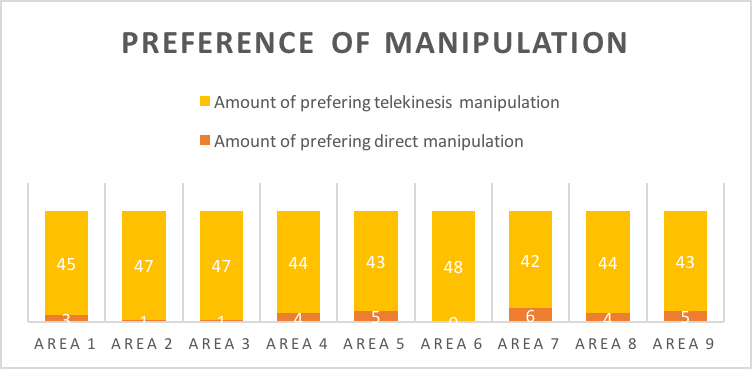}
\caption{Preference of Telekinesis Manipulation}
\label{fig:res2}
\end{figure}

\begin{figure}[h]
\includegraphics[width=0.45\textwidth]{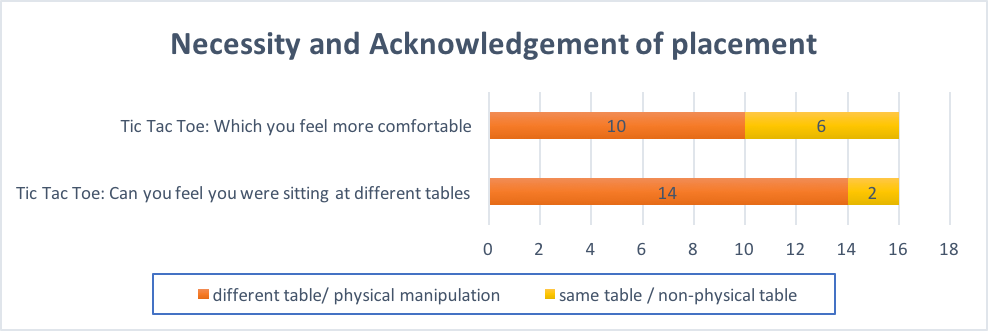}
\caption{Necessity and Acknowledgement of placement}
\label{fig:res3}
\end{figure}

For tiles one, two, and three (far from the player) in Figure 5, forty-three out of forty-eight (89.6\%) were moved using gestures. For tiles four, five, and six (at a medium distance from the player), thirty-nine out of forty-eight (81.3\%) were moved using gestures. For the close tiles, including seven, eight, and nine, thirty-three out of forty-eight (68.8\%) were moved using gestures.

While most participants preferred to use gestures for distant targets and physical interaction for proximate targets, one participant mentioned that they preferred gestures over physical interaction because the gestures were more fun. Two other users both chose to use gestures to push proximate targets away before pulling them back again. We concluded from this that the interaction we designed is not only meaningful and useful, but enjoyable as well.

\section{Limitations and Future Work}

One limitation is our hardware configuration. Currently, repurposing our robots after installation is not easy. Additionally, our robots cannot assume dynamic shapes.
For remote collaboration, we would like to try a more sophisticated task, in order to determine how tasks with varying levels of challenge require different methods of design.

\section{Conclusion}
In this paper, we proposed a new approach for interaction in virtual reality via robotic haptic proxies, specifically targeted towards collaborative experiences, both remote and local. We presented several prototypes utilizing our three mapping definitions, demonstrating that
robotic proxies can be temporarily assigned to represent different virtual objects, that our system can allow remotely located users to have the experience of touching on the same virtual object, and that users can alternately use gestures to command objects without touching them. 
Our preliminary experiments returned positive results. In the future we plan to undertake a more comprehensive study focusing on deeper application interactions and expanded hardware capability.

\balance{}

\bibliographystyle{SIGCHI-Reference-Format}
\bibliography{sample}

\end{document}